\documentclass[aps,prl,twocolumn,superscriptaddress]{revtex4-1}

\usepackage{graphicx} 
\usepackage{epsfig}
\usepackage{amsmath} 
\usepackage{amsthm} 
\usepackage{amssymb}	

\usepackage{graphics} 
\usepackage{hyperref} 
\hypersetup{
    colorlinks,
    citecolor=blue,
    filecolor=black,
    linkcolor=blue,
    urlcolor=blue
}

\begin{document}


\title{Comment on ``Repulsive Contact Interactions Make Jammed Particulate Systems Inherently Nonharmonic"}


\author{Carl P. Goodrich}
\email[]{cpgoodri@sas.upenn.edu}

\author{Andrea J. Liu}
\affiliation{Department of Physics, University of Pennsylvania, Philadelphia, Pennsylvania 19104, USA}

\author{Sidney R. Nagel}
\affiliation{James Franck and Enrico Fermi Institutes, The University of Chicago, Chicago, Illinois 60637, USA}


\begin{abstract}
\end{abstract}


\maketitle
In their Letter~\cite{Schreck:2011kl}, Schreck, Bertrand, O'Hern and Shattuck (SBOS) study nonlinearities in jammed particulate systems that arise when contacts are altered. They conclude that there is ``no harmonic regime in the large system limit ($N\rightarrow \infty$) for all compressions $\Delta \phi$" and that ``the density of vibrational modes cannot be described using the dynamical matrix ..., which dramatically affects mechanical response, specific heat and energy diffusivity." We dispute these conclusions and argue that linear response is justified and essential for understanding the nature of jammed solids, even at nonzero temperature $T$.

Linear response holds if the displacements, $\delta$, can be made small enough so that a quadratic expansion of the energy in terms of $\delta$ provides an accurate description of macroscopic thermal and mechanical properties of the system. 
The distinction between microscopic and macroscopic phenomena is important, as many crystals lack a regime where the dynamical matrix perfectly describes all aspects of the response, except strictly for $\delta=0$ or $T=0$~\cite{Ashcroft:1976ud}.
The argument of SBOS rests on the claim that for ``one-sided repulsive interactions," the changing of a single contact is sufficient to destroy the linear regime because it can cause energy to leak from a mode. However, as is well known for crystals, such failure at the microscopic level does not imply that bulk harmonic properties, such as the density of vibrational modes, elastic constants or heat capacity, fail to predict response and thus does not preclude a linear regime.


Like crystals, jammed solids have a valid linear regime in the thermodynamic limit. Note that there are two important limits: the limit $N\rightarrow \infty$ and the limit $\delta\rightarrow 0$. SBOS agree that above the jamming transition, there is a linear regime for systems with finite $N$. Therefore, if $\delta\rightarrow 0$ before $N\rightarrow \infty$ there is clearly a well-defined linear regime. Moreover, if the limits are reversed ($N \rightarrow \infty$ before $\delta \rightarrow 0$),
the linear regime persists even if an extensive number of contacts are altered. In perturbation theory, changing a single contact affects the density of states by at most $\frac 1N$ ~\cite{Goodrich2013}.  Therefore, we can choose a nonzero $\delta$ so that the cumulative effect of contact changes is negligibly small on the density of states and other bulk harmonic properties.  Thus there is always a well-defined linear regime as $N\rightarrow \infty$, in agreement with Ref.~\cite{Ikeda:2013gu}.

No real system can be described by purely repulsive Hookian springs, the example potential used in the SBOS study. It is therefore important to understand how changes to this potential affect their conclusions.
The SBOS argument fails for any finite-range repulsive potential that does not have a discontinuity in either of its first two derivatives at its cutoff, such as the Hertzian interaction~\cite{Goodrich2013}.  For such interactions, the dynamical matrix is continuous so for $\delta < \delta_0$ (where $\delta_0$ is positive and depends on the potential) there is no appreciable effect on the force. Thus, even if contacts change, the harmonic approximation is valid for $\delta < \delta_0$~\cite{Goodrich2013}.  Only for the special case of Hookian repulsions, where $\delta_0=0$, do we need to invoke the more subtle perturbation argument above to understand that a linear regime exists.

The SBOS claim that ``the density of vibrational modes cannot be described using the dynamical matrix" is inconsistent with the generally accepted definition of ``modes"; 
normal modes of vibration are \emph{by definition} a linear approximation~\cite{Chaikin:2000td}. 
The temperature $T^*$ above which linear behavior breaks down is currently a matter of debate~\cite{Ikeda:2013gu,Wang:2013gy}, but there is no doubt that $T^* >0$ above the jamming transition.
  
Although anharmonic effects dominate at the jamming transition ($\Delta\phi \rightarrow 0$)~\cite{OHern:2003vq,Xu:2010fa}, it is still essential to understand how linear quantities behave as $\Delta\phi \rightarrow 0$ (just as it is important to understand the linear susceptibility near the Ising critical point). 
One also needs linear response to understand nonlinear effects.  For example, energy barriers depend strongly on mode frequency~\cite{Xu:2010fa}, particle rearrangements are strongly correlated with low-frequency modes~\cite{Manning:2011dk}, and shock fronts in marginally jammed solids depend on how the sound speed vanishes~\cite{Gomez:2012ji}.   Linear response thus provides a valid and indispensable foundation for understanding jammed solids.


We thank the U.S. Department of Energy, Office of Basic Energy Sciences, Division of Materials Sciences and Engineering under Awards DE-FG02-05ER46199 and DE-FG02-03ER46088.  CPG was supported by an NSF Graduate Research Fellowship.


\begin{thebibliography}{10}%
\makeatletter
\providecommand \@ifxundefined [1]{%
 \@ifx{#1\undefined}
}%
\providecommand \@ifnum [1]{%
 \ifnum #1\expandafter \@firstoftwo
 \else \expandafter \@secondoftwo
 \fi
}%
\providecommand \@ifx [1]{%
 \ifx #1\expandafter \@firstoftwo
 \else \expandafter \@secondoftwo
 \fi
}%
\providecommand \natexlab [1]{#1}%
\providecommand \enquote  [1]{``#1''}%
\providecommand \bibnamefont  [1]{#1}%
\providecommand \bibfnamefont [1]{#1}%
\providecommand \citenamefont [1]{#1}%
\providecommand \href@noop [0]{\@secondoftwo}%
\providecommand \href [0]{\begingroup \@sanitize@url \@href}%
\providecommand \@href[1]{\@@startlink{#1}\@@href}%
\providecommand \@@href[1]{\endgroup#1\@@endlink}%
\providecommand \@sanitize@url [0]{\catcode `\\12\catcode `\$12\catcode
  `\&12\catcode `\#12\catcode `\^12\catcode `\_12\catcode `\%12\relax}%
\providecommand \@@startlink[1]{}%
\providecommand \@@endlink[0]{}%
\providecommand \url  [0]{\begingroup\@sanitize@url \@url }%
\providecommand \@url [1]{\endgroup\@href {#1}{\urlprefix }}%
\providecommand \urlprefix  [0]{URL }%
\providecommand \Eprint [0]{\href }%
\providecommand \doibase [0]{http://dx.doi.org/}%
\providecommand \selectlanguage [0]{\@gobble}%
\providecommand \bibinfo  [0]{\@secondoftwo}%
\providecommand \bibfield  [0]{\@secondoftwo}%
\providecommand \translation [1]{[#1]}%
\providecommand \BibitemOpen [0]{}%
\providecommand \bibitemStop [0]{}%
\providecommand \bibitemNoStop [0]{.\EOS\space}%
\providecommand \EOS [0]{\spacefactor3000\relax}%
\providecommand \BibitemShut  [1]{\csname bibitem#1\endcsname}%
\let\auto@bib@innerbib\@empty
\bibitem [{\citenamefont {Schreck}\ \emph {et~al.}(2011)\citenamefont
  {Schreck}, \citenamefont {Bertrand}, \citenamefont {O'Hern},\ and\
  \citenamefont {Shattuck}}]{Schreck:2011kl}%
  \BibitemOpen
  \bibfield  {author} {\bibinfo {author} {\bibfnamefont {C.~F.}\ \bibnamefont
  {Schreck}}, \bibinfo {author} {\bibfnamefont {T.}~\bibnamefont {Bertrand}},
  \bibinfo {author} {\bibfnamefont {C.~S.}\ \bibnamefont {O'Hern}}, \ and\
  \bibinfo {author} {\bibfnamefont {M.~D.}\ \bibnamefont {Shattuck}},\
  }\href@noop {} {\bibfield  {journal} {\bibinfo  {journal} {Phys. Rev. Lett.}\
  }\textbf {\bibinfo {volume} {107}},\ \bibinfo {pages} {078301} (\bibinfo
  {year} {2011})}\BibitemShut {NoStop}%
\bibitem [{\citenamefont {Ashcroft}\ and\ \citenamefont
  {Mermin}(1976)}]{Ashcroft:1976ud}%
  \BibitemOpen
  \bibfield  {author} {\bibinfo {author} {\bibfnamefont {N.~W.}\ \bibnamefont
  {Ashcroft}}\ and\ \bibinfo {author} {\bibfnamefont {N.~D.}\ \bibnamefont
  {Mermin}},\ }\href@noop {} {\emph {\bibinfo {title} {{Solid state
  physics}}}}\ (\bibinfo  {publisher} {Thomson Brooks/Cole},\ \bibinfo {year}
  {1976})\BibitemShut {NoStop}%
\bibitem [{\citenamefont {Goodrich}\ \emph {et~al.}(2013)\citenamefont
  {Goodrich}, \citenamefont {Liu},\ and\ \citenamefont {Nagel}}]{Goodrich2013}%
  \BibitemOpen
  \bibfield  {author} {\bibinfo {author} {\bibfnamefont {C.~P.}\ \bibnamefont
  {Goodrich}}, \bibinfo {author} {\bibfnamefont {A.~J.}\ \bibnamefont {Liu}}, \
  and\ \bibinfo {author} {\bibfnamefont {S.~R.}\ \bibnamefont {Nagel}},\
  }\href@noop {} {\bibfield  {journal} {\bibinfo  {journal} {unpublished~}\ }
  (\bibinfo {year} {2013})}\BibitemShut {NoStop}%
\bibitem [{\citenamefont {Ikeda}\ \emph {et~al.}(2013)\citenamefont {Ikeda},
  \citenamefont {Berthier},\ and\ \citenamefont {Biroli}}]{Ikeda:2013gu}%
  \BibitemOpen
  \bibfield  {author} {\bibinfo {author} {\bibfnamefont {A.}~\bibnamefont
  {Ikeda}}, \bibinfo {author} {\bibfnamefont {L.}~\bibnamefont {Berthier}}, \
  and\ \bibinfo {author} {\bibfnamefont {G.}~\bibnamefont {Biroli}},\
  }\href@noop {} {\bibfield  {journal} {\bibinfo  {journal} {J. Chem. Phys.}\
  }\textbf {\bibinfo {volume} {138}},\ \bibinfo {pages} {12A507} (\bibinfo
  {year} {2013})}\BibitemShut {NoStop}%
\bibitem [{\citenamefont {Chaikin}\ and\ \citenamefont
  {Lubensky}(2000)}]{Chaikin:2000td}%
  \BibitemOpen
  \bibfield  {author} {\bibinfo {author} {\bibfnamefont {P.~M.}\ \bibnamefont
  {Chaikin}}\ and\ \bibinfo {author} {\bibfnamefont {T.~C.}\ \bibnamefont
  {Lubensky}},\ }\href@noop {} {\emph {\bibinfo {title} {{Principles of
  Condensed Matter Physics}}}}\ (\bibinfo  {publisher} {Cambridge University
  Press},\ \bibinfo {year} {2000})\BibitemShut {NoStop}%
\bibitem [{\citenamefont {Wang}\ and\ \citenamefont {Xu}(2013)}]{Wang:2013gy}%
  \BibitemOpen
  \bibfield  {author} {\bibinfo {author} {\bibfnamefont {L.}~\bibnamefont
  {Wang}}\ and\ \bibinfo {author} {\bibfnamefont {N.}~\bibnamefont {Xu}},\
  }\href@noop {} {\bibfield  {journal} {\bibinfo  {journal} {Soft Matter}\
  }\textbf {\bibinfo {volume} {9}},\ \bibinfo {pages} {2475} (\bibinfo {year}
  {2013})}\BibitemShut {NoStop}%
\bibitem [{\citenamefont {O'Hern}\ \emph {et~al.}(2003)\citenamefont {O'Hern},
  \citenamefont {Silbert}, \citenamefont {Liu},\ and\ \citenamefont
  {Nagel}}]{OHern:2003vq}%
  \BibitemOpen
  \bibfield  {author} {\bibinfo {author} {\bibfnamefont {C.~S.}\ \bibnamefont
  {O'Hern}}, \bibinfo {author} {\bibfnamefont {L.~E.}\ \bibnamefont {Silbert}},
  \bibinfo {author} {\bibfnamefont {A.~J.}\ \bibnamefont {Liu}}, \ and\
  \bibinfo {author} {\bibfnamefont {S.~R.}\ \bibnamefont {Nagel}},\ }\href@noop
  {} {\bibfield  {journal} {\bibinfo  {journal} {Phys. Rev. E}\ }\textbf
  {\bibinfo {volume} {68}},\ \bibinfo {pages} {011306} (\bibinfo {year}
  {2003})}\BibitemShut {NoStop}%
\bibitem [{\citenamefont {Xu}\ \emph {et~al.}(2010)\citenamefont {Xu},
  \citenamefont {Vitelli}, \citenamefont {Liu},\ and\ \citenamefont
  {Nagel}}]{Xu:2010fa}%
  \BibitemOpen
  \bibfield  {author} {\bibinfo {author} {\bibfnamefont {N.}~\bibnamefont
  {Xu}}, \bibinfo {author} {\bibfnamefont {V.}~\bibnamefont {Vitelli}},
  \bibinfo {author} {\bibfnamefont {A.~J.}\ \bibnamefont {Liu}}, \ and\
  \bibinfo {author} {\bibfnamefont {S.~R.}\ \bibnamefont {Nagel}},\ }\href@noop
  {} {\bibfield  {journal} {\bibinfo  {journal} {EPL}\ }\textbf {\bibinfo
  {volume} {90}},\  (\bibinfo {year} {2010})}\BibitemShut {NoStop}%
\bibitem [{\citenamefont {Manning}\ and\ \citenamefont
  {Liu}(2011)}]{Manning:2011dk}%
  \BibitemOpen
  \bibfield  {author} {\bibinfo {author} {\bibfnamefont {M.~L.}\ \bibnamefont
  {Manning}}\ and\ \bibinfo {author} {\bibfnamefont {A.~J.}\ \bibnamefont
  {Liu}},\ }\href@noop {} {\bibfield  {journal} {\bibinfo  {journal} {Phys.
  Rev. Lett.}\ }\textbf {\bibinfo {volume} {107}},\ \bibinfo {pages} {108302}
  (\bibinfo {year} {2011})}\BibitemShut {NoStop}%
\bibitem [{\citenamefont {G{\'o}mez}\ \emph {et~al.}(2012)\citenamefont
  {G{\'o}mez}, \citenamefont {Turner}, \citenamefont {van Hecke},\ and\
  \citenamefont {Vitelli}}]{Gomez:2012ji}%
  \BibitemOpen
  \bibfield  {author} {\bibinfo {author} {\bibfnamefont {L.~R.}~\bibnamefont
  {G{\'o}mez}}, \bibinfo {author} {\bibfnamefont {A.~M.}~\bibnamefont {Turner}},
  \bibinfo {author} {\bibfnamefont {M.}~\bibnamefont {van Hecke}}, \ and\
  \bibinfo {author} {\bibfnamefont {V.}~\bibnamefont {Vitelli}},\ }\href@noop
  {} {\bibfield  {journal} {\bibinfo  {journal} {Phys. Rev. Lett.}\ }\textbf
  {\bibinfo {volume} {108}},\ \bibinfo {pages} {058001} (\bibinfo {year}
  {2012})}\BibitemShut {NoStop}%
\end{thebibliography}

%

\end{document}